\pdfoutput=1
\PassOptionsToPackage{table}{xcolor}
\documentclass[sigconf, nonacm, 10pt]{acmart}
\AtBeginDocument{%
	}
		
\usepackage{mathrsfs, amsmath,amsfonts}
\usepackage{graphicx}
\usepackage{textcomp}
\usepackage{mdframed}
\usepackage[table]{xcolor}

\usepackage{lipsum}

\usepackage[utf8]{inputenc}
\usepackage[T1]{fontenc}

\usepackage[abbreviations]{foreign}
\usepackage{algorithm}
\usepackage{algpseudocode}

\usepackage[labelfont=bf,labelsep=period]{caption}
\usepackage[nodayofweek]{datetime}
\usepackage[inline]{enumitem}
\usepackage{float}
\usepackage{geometry}
\usepackage{graphicx}
\usepackage[utf8]{inputenc}
\usepackage{times}
\usepackage{url}

\usepackage{xspace}
\usepackage{pifont}

\newdateformat{simple}{\THEDAY\ \monthname[\THEMONTH]\ \THEYEAR}
\simple

\floatstyle{ruled}
\newfloat{algo}{htbp}{algo}
\floatname{algo}{Algorithm}

\def \ifempty#1{\def\temp{#1} \ifx\temp\empty }

    \usepackage{textcomp}
    \usepackage{listings}
    \usepackage{booktabs} 
    \usepackage{multirow}
    \usepackage{tabularx}
    \usepackage{makecell}
    \usepackage{float}
    \usepackage[export]{adjustbox}
    \usepackage{eso-pic}
    \usepackage[most]{tcolorbox}
    \usepackage{calc}
    
    \usepackage{needspace}

\newcommand{\yes}{\textcolor{green}{\ding{51}}}
\newcommand{\no}{\textcolor{red}{\ding{55}}}

    \renewcommand{\arraystretch}{1}

    \newcolumntype{R}{>{\raggedleft\arraybackslash}X}
    \newcolumntype{P}[1]{>{\raggedleft\arraybackslash}p{#1}}

    \definecolor{prioritycolor}{HTML}{969bce}
    \definecolor{darkgray}{HTML}{262626}

    \usepackage{fontawesome5}
    \newboolean{showcomments}
    \setboolean{showcomments}{true}
    \ifthenelse{\boolean{showcomments}}
    { \newcommand{\mynote}[3]{
            \fbox{\bfseries\sffamily\scriptsize#1}
    {\small$\blacktriangleright$\textsf{\emph{\color{#3}{#2}}}$\blacktriangleleft$}}}
    { \newcommand{\mynote}[3]{}}

    \newcounter{numobserv} 
    \definecolor{beaublue}{rgb}{0.88, 0.93, 0.93}

    \usepackage[most]{tcolorbox}
    \colorlet{shadecolor}{beaublue}
    \tcbset{
        colback=beaublue,
        boxrule=0.5pt,
        boxsep=0pt,
        left=4pt,
        right=4pt,
        top=4pt,
        bottom=4pt,
        sharp corners,
        colframe=gray!70,
    }
    \newcommand{\observ}[1]{
        \addtocounter{numobserv}{1}
        \begin{tcolorbox}	
            \textit{\textbf{Take-away\,\thenumobserv\,:} #1 }	
        \end{tcolorbox}
    }

    \newtcolorbox{variantbox}[1]{%
        enhanced, 
        attach boxed title to top center={yshift=-2.8mm,yshifttext=-1mm,xshift=-10mm},
        colback=gray!10, 
        colframe=black, 
        colbacktitle=black, 
        sharp corners,
        top=10pt,
        fonttitle=\color{black},
        title=\textbf{#1},   
        fontupper= \ttfamily,
        boxed title style={size=small, colframe=black, colback=gray!10, sharp corners, boxrule=0.5pt} 
    }

    \usepackage{titlesec}
    \renewcommand{\thesubsubsection}{\arabic{subsubsection}.}
    \titleformat{\subsubsection}[runin]
    {\normalfont\bfseries\itshape}
{\thesubsubsection}{0.5em}{}[\hspace{0.5em}\\~\\]

\newcommand{\subpoint}[1]{\smallskip\noindent\textbf{#1}\xspace}

\newcommand{\setup}{\ensuremath{\mathsf{Setup}}}
\newcommand{\evl}{\ensuremath{\mathsf{Eval}}}

\newcommand{\tfheenc}{\ensuremath{\mathsf{ThFHE}}\text{-}\ensuremath{\mathsf{Enc}}}
\newcommand{\tfhedec}{\ensuremath{\mathsf{ThFHE}}\text{-}\ensuremath{\mathsf{Dec}}}
\newcommand{\tfheeval}{\ensuremath{\mathsf{ThFHE}}\text{-}\ensuremath{\mathsf{Eval}}}
\newcommand{\tfhecomb}{\ensuremath{\mathsf{ThFHE}}\text{-}\ensuremath{\mathsf{Combine}}}
\newcommand{\pkenc}{\ensuremath{\mathsf{PK}}\text{-}\ensuremath{\mathsf{Enc}}}
\newcommand{\teeskdec}{\ensuremath{\mathsf{TEE}}\text{-}\ensuremath{\mathsf{SK}}\text{-}\ensuremath{\mathsf{Dec}}}
\newcommand{\teeeval}{\ensuremath{\mathsf{TEE}}\text{-}\ensuremath{\mathsf{Eval}}}
\newcommand{\teeagg}{\ensuremath{\mathsf{TEE}}\text{-}\ensuremath{\mathsf{Aggregate}}}
\newcommand{\ot}{\ensuremath{\mathsf{OT}}}

\newcommand{\snd}{\ensuremath{\mathsf{Send}}}

\newcommand{\varone}{\((P_\mathsf{NT},\allowbreak A_\mathsf{NT},\allowbreak Q_\mathsf{NC})\)\xspace}
\newcommand{\vartwo}{\((P_\mathsf{NT},\allowbreak A_\mathsf{T},\allowbreak Q_\mathsf{NC})\)\xspace}
\newcommand{\varthree}{\((P_\mathsf{NT},\allowbreak A_\mathsf{NT},\allowbreak Q_\mathsf{C})\)\xspace}
\newcommand{\varfour}{\((P_\mathsf{NT},\allowbreak A_\mathsf{T},\allowbreak Q_\mathsf{C})\)\xspace}
\newcommand{\varfive}{\((P_\mathsf{T},\allowbreak A_\mathsf{NT},\allowbreak Q_\mathsf{C})\)\xspace}
\newcommand{\varsix}{\((P_\mathsf{T},\allowbreak A_\mathsf{T},\allowbreak Q_\mathsf{C})\)\xspace}

\usepackage{hyperref}
\usepackage{xspace}
\newcommand{\copyrighttext}{  \scriptsize \textcopyright 2025 ACM.
	Personal use of this material is permitted.
	Permission from ACM must be obtained for all other uses,
	in any current or future media, including reprinting/republishing this
	material for advertising or promotional purposes, creating new collective
	works, for resale or redistribution to servers or
	lists, or reuse of any copyrighted component of this work in other works.
	This is the author's pre-print version of the work. Published in the 19th ACM International Conference on Distributed and Event-Based Systems (DEBS).
}

\begin{document}

\title{Practical Secure Aggregation by Combining Cryptography and Trusted Execution Environments}
\renewcommand{\shorttitle}{Practical Secure Aggregation by Combining Cryptography and TEEs}

\author{Romain de Laage}
\affiliation{
	\institution{University of Neuch\^{a}tel}
	\country{Switzerland}}
\email{romain.delaage@unine.ch}
\orcid{0009-0009-3476-1688}

\author{Peterson Yuhala}
\affiliation{
	\institution{University of Neuch\^{a}tel}
	\country{Switzerland}}
\email{peterson.yuhala@unine.ch}
\orcid{0000-0002-3371-9228}

\author{François-Xavier Wicht}
\affiliation{
	\institution{University of Bern}
	\country{Switzerland}}
\email{francois-xavier.wicht@unibe.ch}
\orcid{0009-0005-6090-7901}

\author{Pascal Felber}
\affiliation{
	\institution{University of Neuch\^{a}tel}
	\country{Switzerland}}
\email{pascal.felber@unine.ch}
\orcid{0000-0003-1574-6721}

\author{Christian Cachin}
\affiliation{
	\institution{University of Bern}
	\country{Switzerland}}
\email{christian.cachin@unibe.ch}
\orcid{0000-0001-8967-9213}

\author{Valerio Schiavoni}
\affiliation{
	\institution{University of Neuch\^{a}tel}
	\country{Switzerland}}
\email{valerio.schiavoni@unine.ch}
\orcid{0000-0003-1493-6603}

\renewcommand{\shortauthors}{de Laage et al.}

\date{\today}

\begin{abstract}
Secure aggregation enables a group of mutually distrustful parties, each holding private inputs, to collaboratively compute an aggregate value while preserving the privacy of their individual inputs.
However, a major challenge in adopting secure aggregation approaches for practical applications is the significant computational overhead of the underlying cryptographic protocols, \eg fully homomorphic encryption.
This overhead makes secure aggregation protocols impractical, especially for large datasets.
In contrast, hardware-based security techniques such as trusted execution environments (TEEs) enable computation at near-native speeds, making them a promising alternative for reducing the computational burden typically associated with purely cryptographic techniques.
Yet, in many scenarios, parties may opt for either cryptographic or hardware-based security mechanisms, highlighting the need for hybrid approaches. 
In this work, we introduce several secure aggregation architectures that integrate both cryptographic and TEE-based techniques, analyzing the trade-offs between security and performance.

\end{abstract}

\newcommand{\copyrightnotice}{\begin{tikzpicture}[remember picture,overlay]
	\node[anchor=south,yshift=2pt,fill=yellow!20] at (current page.south) {\fbox{\parbox{\dimexpr\textwidth-\fboxsep-\fboxrule\relax}{\copyrighttext}}};
	\end{tikzpicture}
}
\maketitle
\copyrightnotice

\section{Introduction}
\label{sec:intro}
Modern organizations, including companies and government agencies, regularly need third-party data to guide their decision making, research, or product development~\cite{canetti01}.
However, data privacy considerations and regulations, \eg GDPR~\cite{GDPR2016}, CCPA~\cite{CCPA2018}, HIPAA~\cite{HIPAA1996} increasingly restrict such data sharing.
For instance, consider a health insurance company that collaborates with multiple healthcare providers to gather essential data about a certain disease, and compute the average of infected people.
One possible way is for the insurance provider to request this average from each individual healthcare provider and then aggregate this locally.
However, this approach leaks sensitive information: the average number of infected people per provider, \ie, the data inputs; and the disease that the insurance company wants to inquire about, \ie, the query.
\emph{Secure aggregation} is a cryptographic protocol that allows such multi-party aggregates to be computed while preserving input privacy.
There have been numerous studies showcasing applications of secure aggregation in areas like federated learning~\cite{bonawitz2017practical}, secure e-voting~\cite{kusters2020ordinos}, privacy-preserving auctions~\cite{aly17}, and privacy-preserving data-analytics in general~\cite{danezis13, tariq14, sia}. 

Many secure aggregation schemes rely on purely cryptographic techniques like fully-homomorphic encryption (FHE)~\cite{mpc-tfhe} to perform computations on encrypted data.
While these techniques provide strong privacy guarantees, they incur large computational overhead.
Alternatively, \emph{trusted execution environments} (TEEs) provide a more pragmatic alternative to mitigate these challenges and computational costs.
By leveraging hardware based cryptographic mechanisms, TEEs enable computations on plaintext data within secure enclaves, delivering substantial performance enhancements.
However, their adoption can be constrained by factors such as hardware availability or organizational policies that limit reliance on proprietary TEE implementations due to trust concerns. As a result, privacy-preserving computations like secure aggregation require hybrid architectures that seamlessly integrate TEEs with cryptographic techniques, ensuring both flexibility and robust security guarantees.

Several research work have explored secure aggregation protocols based on purely cryptographic mechanisms~\cite{bonawitz2017practical, hiccups}.
Others have examined outsourcing computations to TEEs to alleviate the computational burden of purely cryptographic approaches~\cite{wu2022hybrid, dantonio2023}.
Our work conducts a thorough investigation of previously unexplored secure aggregation architectures in which parties use either purely cryptographic techniques or hardware-based TEEs, and analyzes the performance-security trade-offs.
In summary, our contributions are as follows:

\begin{enumerate}
	\item A comprehensive exploration of various combinations of hardware- and software-based architectures for secure aggregation.
	\item An in-depth evaluation of said combinations with varying numbers of parties and input data sizes.
	\item A detailed analysis of the performance and security trade-offs across the different approaches.
\end{enumerate}

\section{Background}
\label{sec:background}
\subsection{Secure aggregation}
\emph{Secure aggregation} (SA) is a type of secure multi-party computation (MPC)~\cite{evans2018pragmatic, stevens2022secret} which enables a group of mutually distrustful parties to aggregate their private inputs (using a central aggregator or not), revealing only the aggregate value. 
Secure aggregation protocols have applications in areas like privacy-preserving machine learning~\cite{bonawitz2017practical}, medical research~\cite{hiccups}, and secure data analytics as a whole~\cite{danezis13, tariq14, sia}.
These protocols leverage techniques like fully homomorphic encryption (FHE) or secret sharing to ensure data privacy.

\subpoint{Problem definition.}
We consider a secure aggregation scheme where one party, an \emph{aggregator} $A$ computes a query $q$ (could be confidential or not) which it sends to $n$ parties $P_1,\ldots,P_n$, \ie data providers.
Each party $P_i$ possesses private input $x_i$, and evaluates the query on its private data via a function $f$ to produce a subresult $r_i$, which is encrypted yielding $\hat{r_i}$. 
The encrypted subresults (or at least a given threshold $t$) are then sent to $A$, whose goal is to compute the aggregate $\alpha = \mathit{Aggr}(r_1,\ldots,r_n)$ correctly, while preserving the privacy of the parties' inputs.
To ensure input and query privacy, the parties and aggregator can opt for either a purely cryptographic technique like FHE or hardware-based technique like a TEE.
Our work provides a thorough exploration of secure aggregation architectures to achieve this.

\subsection{Fully homomorphic encryption}
Fully homomorphic encryption (FHE) allows an arbitrary function $f$ to be evaluated over encrypted data~\cite{gentry2009, gentry2010, openfhe}.
FHE provides a method for secure computation between two parties ($P_1$ and $P_2$) where $P_1$ encrypts their data with their public key, and $P_2$ homomorphically evaluates the function on $P_1$'s encrypted data along with their own input~\cite{mpc-tfhe}.
When extending this method to multiple parties, a challenge arises regarding which encryption key to use.
If each party uses a separate public key, homomorphic evaluation on the different ciphertexts becomes infeasible.
Conversely, if a single party selects the key for everyone, compromising this party would jeopardize the privacy of every participant.
\emph{Threshold fully homomorphic encryption (ThFHE)}~\cite{thfhe,mpc-tfhe} alleviates this issue by supporting a $t$-out-of-$n$ threshold decryption protocol.
The common public key can be used to perform HE operations, and a threshold of the parties can collaboratively decrypt and combine the partial results using the shared secret key.

\subpoint{Secret sharing.}
Secret sharing~\cite{shamir79} is a primitive at the heart of many MPC protocols.
Informally, a \((t, n)\)-secret sharing scheme divides a secret \(s\) into \(n\) shares such that any set of at most \(t - 1\) shares provides no information about \(s\), while any set of \(t\) shares enables full reconstruction of the secret \(s\).

\subpoint{Oblivious transfer.}
Oblivious transfer (OT) is a two-party protocol where a sender holds a list of $k$ values $\{x_{[k]}\}$ and the receiver an index $i\in[k]$. 
The protocol allows the receiver to learn $x_i$ without learning anything about $x_j$ with $j\neq i$, while the sender does not learn $i$.

\subsection{Trusted execution environments}
A trusted execution environment is an isolated processing environment provided by the CPU ensuring confidentiality and integrity for code and data during execution.
TEEs are designed to provide strong security guarantees when the adversary has control over the hardware (\eg DRAM) and privileged system software, \ie OS and hypervisor.
The primary advantage of TEEs over purely cryptographic techniques like FHE is that sensitive data is kept encrypted in memory but is transparently decrypted in the CPU, enabling computations to be performed securely on the plaintext data at native speeds, as opposed to FHE which keeps the data encrypted at all times.
Popular examples of TEE technologies include Intel SGX~\cite{vcostan} and Arm TrustZone~\cite{pinto19} which provide process-based isolation, and Intel TDX~\cite{cheng2023intel}, AMD SEV~\cite{sev-whitepaper}, and Arm CCA~\cite{arm-cca} which provide virtual machine (VM)-based isolation.
We focus on Intel SGX, which is the most popular process-based TEE technology deployed in cloud infrastructures.
Intel SGX enables applications to create secure memory regions called \emph{enclaves}.
Enclave memory pages are backed by an encrypted region of DRAM called the \emph{enclave page cache} (EPC).
EPC data is transparently decrypted within the CPU package when it is loaded from the EPC into a cache line, and encrypted prior to being written to DRAM.
To facilitate deployment of legacy applications in SGX enclaves, library operating systems (\ie LibOSes) have been developed.
Popular examples include Gramine~\cite{gramine} and Occlum~\cite{occlum}.

\section{Adversarial model}
\label{sec:threat-model}
We assume a \textbf{semi-honest (honest but curious)} model where all parties are guaranteed to follow the protocol specification (\eg they do not submit inaccurate data), but may try to extract information regarding other parties' private data from the messages they receive, \eg a data provider trying to uncover details of a confidential query sent by $A$, or the latter trying to learn private information from the subresult obtained from a data provider. 

\subpoint{Key distribution.}
We assume the existence of a public-key infrastructure as well as cryptographic primitives that make secure encrypted communication channels possible, ensuring the confidentiality and integrity of messages exchanged between the parties.
The key exchanges are performed during a setup phase at initialization, and need not be repeated during periodic aggregation rounds.

\subpoint{TEE security model.} In line with the semi-honest model, a TEE-enabled party may wish to learn private information of another party.
They may proceed by observing system memory or probing the memory channel, but this only reveals encrypted information.
The underlying OS or hypervisor is also under their complete control.
However, they do not carry out operations which may corrupt computational results for example, as this violates the semi-honest adversarial model.
Similarly, we assume that the party is unable to physically open and manipulate the on-premises processor packages.
We assume that the TEE implementations used provide robust remote attestation protocols (\eg Intel SGX) which allow to verify the integrity of code and authenticate the CPU package.
Parties can perform this attestation step during a setup phase.
TEE-based side-channel attacks~\cite{kocher2020spectre, meltdown} for which mitigations exist~\cite{liu16, gruss2017kaslr} are considered out-of-scope.

\section{Architecture}
\label{sec:arch}
This section describes the hybrid secure aggregation architectures/variants considered.

\subpoint{Notation.}
In the remainder of this section, we adopt the following notations: given a plaintext value $a$, the corresponding encrypted value/ciphertext is denoted as $\hat{a}$.
$[n]$ represents the set of positive integers $\{1,\ldots, n\}$ and $\{v_{[n]}\}$ represents the set of values $\{v_1,\ldots,v_n\}$.

\begin{figure}[h!]
	\centering
	\includegraphics[width=0.45\textwidth]{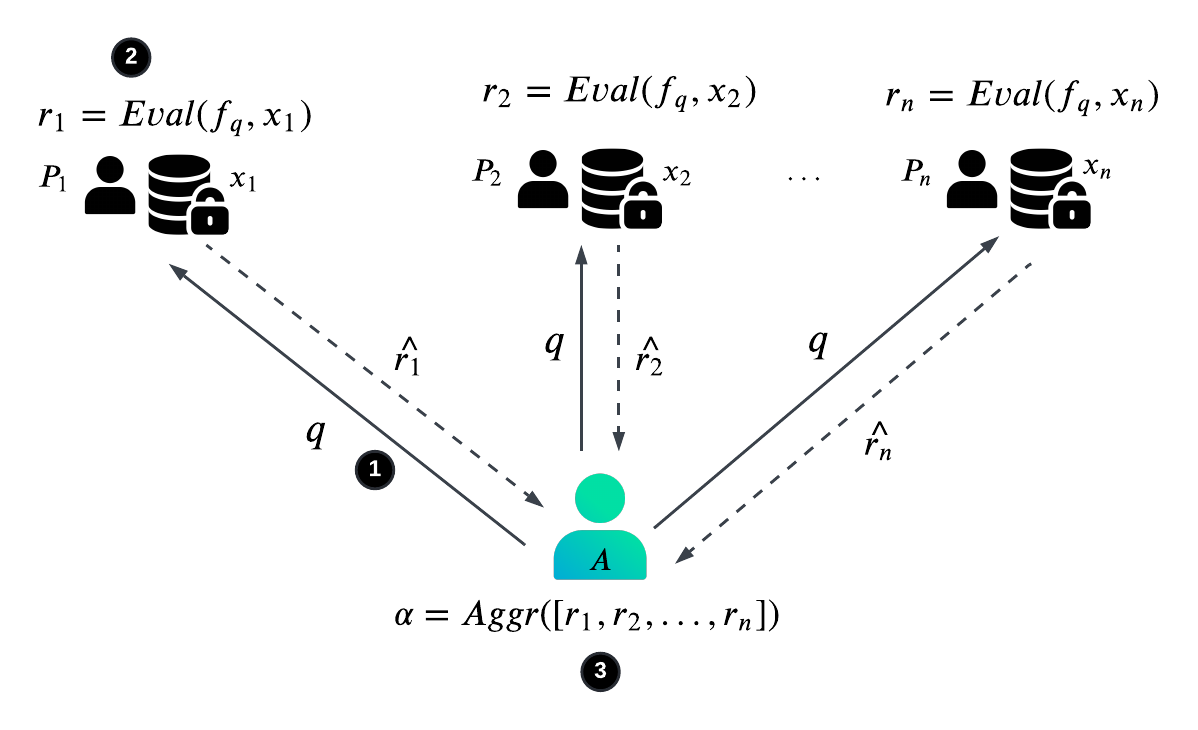}
	\caption{An overview of the generic secure aggregation architecture considered. We explore different variants of this architecture where the parties and aggregator adopt different mechanisms for ensuring data privacy.}
	\label{fig:arch}
\end{figure}

\subsection{Protocol definition}
\autoref{fig:arch} represents a high-level overview of the active phases of the general secure aggregation protocol considered; the protocol execution proceeds in four phases, an offline phase: \emph{setup}, and three online phases: \emph{query dispatch}, \emph{query evaluation}, and \emph{results aggregation}.

\smallskip\noindent\textbf{\emph{Setup}.}
The aim of this phase is to initialize the secure computation protocol. It provides the public parameters $pp$, which are taken as implicit input to the secure computation algorithms.
It comprises public-private key pair generation by all parties for encryption and decryption, as well as other shared keys used for cryptographic primitives like oblivious transfer.
These keys include those used for FHE/ThFHE (a secret key shared by each party and a joint public key) as well as a Curve25519 keypair per node.
The data providers then establish secure communication channels with the aggregator to exchange encrypted messages.
This phase also includes remote attestation of TEE code at the parties and aggregator, as well as authentication of the TEE-enabled platforms.

\smallskip\noindent\textbf{\emph{Query dispatch}.}
In this phase (\autoref{fig:arch} \ding{202}) the aggregator sends the query to each data provider $P_i$; the query could be a summation/frequency count, comparison (min/max), average, set intersection, \etc.
For confidential queries, the aggregator encrypts the query prior to sending.
The encryption key used for this task varies depending on the security tools used by the party.
For example, for a TEE-enabled data provider, the enclave's public key is used.
On the other hand, if there is no TEE at the data provider, a shared key between the aggregator and each data provider may be used to obtain the desired encrypted query result from the data provider via oblivious transfer.

\smallskip\noindent\textbf{\emph{Query evaluation}.}
In this phase (\autoref{fig:arch} \ding{203}), each data provider $P_i$ evaluates $q$ on their private data $x_i$ to obtain a subresult $r_i$.
In our secure aggregation constructions, we refer to this evaluation function as $f$.
The subresult $r_i$ from each $P_i$ is then encrypted to obtain $\hat{r_i}$ which is sent to $A$ for aggregation.
This phase can be run either in a hardware-based TEE, \eg Intel SGX enclave, or following a purely cryptographic approach like oblivious transfer (OT).

\smallskip\noindent\textbf{\emph{Results aggregation}.}
This is the final phase (\autoref{fig:arch} \ding{204}) of the protocol and involves aggregating the subresults received from each $P_i$ via a known aggregation function.
For a TEE-enabled aggregator, the encrypted subresults are first decrypted securely inside the TEE and the aggregation performed on the plaintext subresults.
Otherwise, fully homomorphic encryption is leveraged to obtain the aggregate on the encrypted subresults.
The final result $\alpha$ is then decrypted and shared among all parties via ThFHE.

\subsection{Description of primitives used}
\noindent\textbf{\emph{Primitives used for ThFHE based operations:}}
\begin{itemize}
	\item $\hat{d}=\tfheenc(d, pk)$: encrypts the plaintext $d$ using ThFHE and the joint public key $pk$ as input and returns ciphertext $\hat{d}$
	\item $\hat{\alpha}=\tfheeval(pk, f, \{\hat{r}_{[n]}\})$: evaluates the ThFHE aggregation over the ciphertexts. It takes as input a public key $pk$, a function $f$ represented as a boolean circuit, encrypted data $\hat{r}_i$, and returns an encrypted aggregate $\hat{\alpha}$
	\item $\alpha_i=\tfhedec(\hat{\alpha}, sk_{P_i})$: computes the ThFHE partial decryption of $\hat{\alpha}$ using the secret key share $sk_{P_i}$ and returns a partial decrypted aggregate $\alpha_i$
	\item $\alpha=\tfhecomb(\{\alpha_{[n]}\})$: combines the ThFHE partial decryption of the aggregate $\alpha_i$ to get the plaintext and returns the aggregate $\alpha$
\end{itemize}
\noindent\textbf{\emph{Primitives used for TEE based operations:}}
\begin{itemize}
	\item $\hat{d}=\pkenc\big(d, pk\big)$: encrypts the plaintext $d$ using the public key $pk$ of the TEE and returns the ciphertext $\hat{d}$; can be done within or outside of a TEE
	\item $d=\teeskdec\big(\hat{d},sk\big)$: decrypts $\hat{d}$ using the private key of the TEE $sk$ and returns $d$; must always be done within the TEE
	\item $r=\teeeval(f_q, x)$: evaluates the function $f_q$ associated with the query $q$ within the TEE using the input $x$ and returns the response $r$
	\item $\alpha=\teeagg(\{x_{[n]}\})$: aggregates the values $x_i$ within the TEE and returns the aggregate $\alpha$
\end{itemize}
\noindent\textbf{\emph{Primitives used for OT based operations:}}
\begin{itemize}
	\item $x_i=\ot\big(q, \{x_{[k]}\}\big)$: executes an oblivious transfer between the aggregator and a party. The receiver (aggregator) initiates the transfer to receive the subresult $x_i$ corresponding to $q$ from the sender (party) without revealing the value of $q$. The sender reveals nothing else than the subresult corresponding to $q$.
\end{itemize}

\subsection{Secure aggregation variants}

As mentioned previously, we consider a system with $n$ data providers $P = \{P_1, P_2, \ldots, P_n\}$ and an aggregator $A$. 
These parties may opt for either purely cryptographic techniques or TEEs to ensure data privacy.
We represent the case where parties have TEE capabilities as $P_\mathsf{T}$ and $P_\mathsf{NT}$ otherwise.
Similarly, we define the case where the aggregator has access to trusted hardware as $A_\mathsf{T}$ and $A_\mathsf{NT}$ otherwise.
This results in four configurations: $\{(P_\mathsf{T}A_\mathsf{T}),\allowbreak (P_\mathsf{T}A_\mathsf{NT}),\allowbreak (P_\mathsf{NT}A_\mathsf{T}),\allowbreak (P_\mathsf{NT}A_\mathsf{NT})\}$.
Additionally, the query sent by $A$ could be confidential (denoted as $Q_\mathsf{C}$) or not ($Q_\mathsf{NC}$).
By combining these two possibilities for query confidentiality with the set of TEE configurations for the parties and aggregator, we obtain eight variants. We only consider six of the eight variants, since TEE at parties without confidential queries ($(P_{T}, A_{NT}, Q_{NC})$ and $(P_{T}, A_{T}, Q_{NC})$) brings no useful benefit.
In the rest of this section, we discuss the resulting architectures and security implications of these variants, henceforth represented as $V_i$.
	\subsubsection{No TEE at P, no TEE at A, non-confidential query}
	\label{sec:defvar1}

	\begin{variantbox}{Variant 1:  \( (P_\mathsf{NT}, A_\mathsf{NT}, Q_\mathsf{NC}) \)}
		\textbf{Protocol:}
		\begin{enumerate}[label=\arabic*]
			\item $pp=\setup()$
			\item $A$: $\snd_{A\rightarrow P_i}(q)$
			\item $P_i$: $r_i =\evl(f_q,x_i)$
			\item $P_i$: $\hat{r_i} = \tfheenc(r_i, pk)$
			\item $P_i$: $\snd_{P_i\rightarrow A}(\hat{r_i})$
			\item $A$: $\hat{\alpha} = \tfheeval(pk, f, \{\hat{r}_{[n]}\})$
			\item $A$: $\snd_{A\rightarrow P_i}(\hat{\alpha})$
			\item $P_i$: $\alpha_i = \tfhedec(\hat{\alpha}, {sk_{P_i}})$
			\item $P_i$: $\snd_{P_i\rightarrow A}(\alpha_i)$
			\item $A$: $\alpha=\tfhecomb(\{\alpha_{[n]}\})$
		\end{enumerate}
	\end{variantbox}

\begin{figure}[htbp!]
		\centering
		\includegraphics[width=0.4\textwidth]{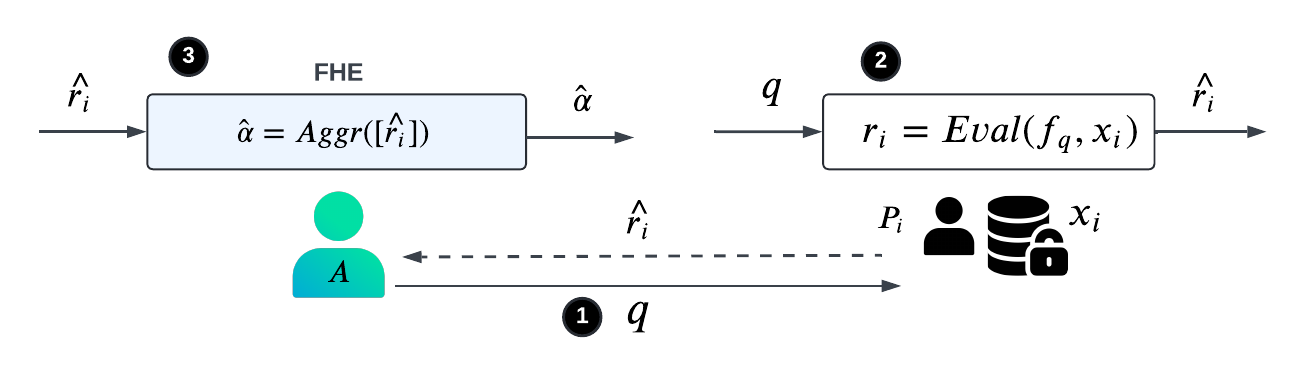}
		\caption{No TEE at P, no TEE at A, non-confidential query}
		\label{fig:arch-v1}
	\end{figure}

	In this variant (described in \autoref{fig:arch-v1}), an unencrypted (thus non-confidential) query is sent by $A$ to all parties (Variant 1: line 2).
	Since $q$ is not confidential, each party simply executes $q$ over its private data to obtain a subresult $r_i$ ($V_1$: line 3).
	Each party then encrypts its subresults $r_i$, yielding $\hat{r_i}$, which is then sent to the aggregator ($V_1$: lines 4-5).
	The latter then aggregates the subresults via FHE, which maintains the privacy of the subresults, and hence parties' sensitive data ($V_1$: line 6).
	The final aggregate value is decrypted via ThFHE ($V_1$: lines 7-10).
	In this variant, there is no reliance on hardware-assisted security, \ie TEEs; the privacy guarantees are provided entirely by software-based cryptographic techniques like FHE.

	\subsubsection{No TEE at P, TEE at A, non-confidential query}
	\label{sec:defvar2}

	\begin{variantbox}{Variant 2: \( (P_\mathsf{NT}, A_\mathsf{T}, Q_\mathsf{NC}) \)}
		\textbf{Protocol:}
		\begin{enumerate}[label=\arabic*]
			\item $pp=\setup()$
			\item $A$: $\snd_{A\rightarrow P_i}(q)$
			\item $P_i$: $r_i = \evl(f_q,x_i)$
			\item $P_i$: $\hat{r}_i = \pkenc\big(r_i, pk_A\big)$
			\item $P_i$: $\snd_{P_i\rightarrow A}(\hat{r}_i)$
			\item $A$: $\{r_{[n]}\} = \teeskdec\big(\{\hat{r}_{[n]}\},sk_A\big)$
			\item $A$: $\alpha = \teeagg\big(\{r_{[n]}\}\big)$
		\end{enumerate}
	\end{variantbox}

\begin{figure}[htbp!]
		\centering
		\includegraphics[width=0.4\textwidth]{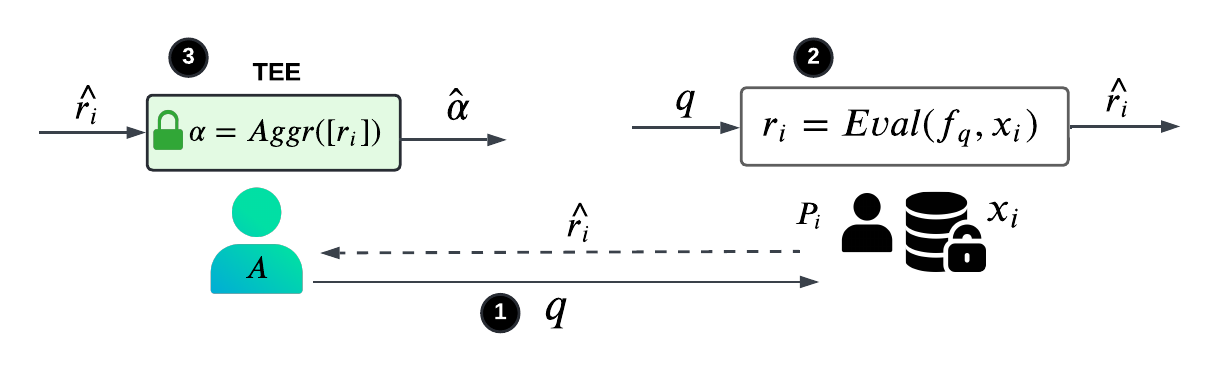}
		\caption{No TEE at P, TEE at A, non-confidential query}
		\label{fig:arch-v2}
	\end{figure}

	Similar to \varone, the parties simply execute $q$ in the clear to obtain subresults $r_i$ which are then encrypted and sent to $A$.
	However, as described in \autoref{fig:arch-v2}, since the aggregator is equipped with a TEE, the subresults can be securely decrypted and aggregated within the TEE to obtain the final aggregate value $\alpha$.\footnote{We note that all TEE-based primitives and their results, \eg \teeskdec, \teeagg~ are done within a TEE and the data being processed at runtime is not accessible to the party or aggregator.}
	We recall that, in a purely cryptographic context, ThFHE is required to achieve threshold decryption.
	That is, the final aggregate is only obtained when a specific threshold, $t$, of subresults has been received by the aggregator.
	However, when the aggregator is equipped with a TEE, a purely crypto-based threshold decryption algorithm is not needed in the TEE.
	The TEE aggregation algorithm can be implemented to enforce the requirement for an aggregate to be performed only after $t$ subresults have been received.
	Such an algorithm is depicted in \autoref{algo:tee-threshold}.
	During the setup phase, all parties can verify this code as part of the remote attestation process.

	\begin{algorithm}[htbp!]
		\caption{TEE-based threshold decryption of subresults and aggregation}
		\label{algo:tee-threshold}
		\begin{algorithmic}[1]
			\State \textbf{Input:} List of encrypted subresults $\{\hat{r_i}\}$, threshold $t$
			\State \textbf{Output:} Decrypted aggregate result $\alpha$ or error
			\State $\mathsf{count} = \mathsf{length}(\{\hat{r}_i\})$
			\If{$count < t$}
			\State \Return \textbf{Error: Threshold not reached}
			\Else

			\State $\{r_i\} = \teeskdec(\{\hat{r}_i\})$
			\State $\alpha =\teeagg(\{r_i\})$
			\State \Return Aggregated result: $\alpha$
			\EndIf
		\end{algorithmic}
	\end{algorithm}

	\needspace{2\baselineskip} 

	\subsubsection{No TEE at P, no TEE at A, confidential query}
	\label{sec:defvar3}

	\begin{variantbox}{Variant 3: \( (P_\mathsf{NT}, A_\mathsf{NT}, Q_\mathsf{C}) \)}
		\textbf{Protocol:}
		\begin{enumerate}[label=\arabic*]
			\item $pp=\setup()$
			\item $P_i$: $\{r_{i_{[k]}}\}=\evl\big(\{f_{q_{[k]}}\},x_i\big)$
			\item $P_i$: $\{\hat{r}_{i_{[k]}}\}=\tfheenc\big(\{r_{i_{[k]}}\},pk\big)$
			\item $A,P_i$: $\hat{r}_{i}=\ot\big(q, \{\hat{r}_{i_{[k]}}\}\big)$
			\item $A$: $\hat{\alpha} = \tfheeval(pk, f, \{\hat{r}_{[n]}\}$)
			\item $A$: $\snd_{A\rightarrow P_i}$($\hat{\alpha}$)
			\item $P_i$: $\alpha_i = \tfhedec(\hat{\alpha}, {sk_{P_i}})$
			\item $P_i$: $\snd_{P_i\rightarrow A}$($\alpha_i$)
			\item $A$: $\alpha=\tfhecomb(\{\alpha_{[n]}\})$
		\end{enumerate}
	\end{variantbox}

\begin{figure}[htbp!]
		\centering
		\includegraphics[width=0.4\textwidth]{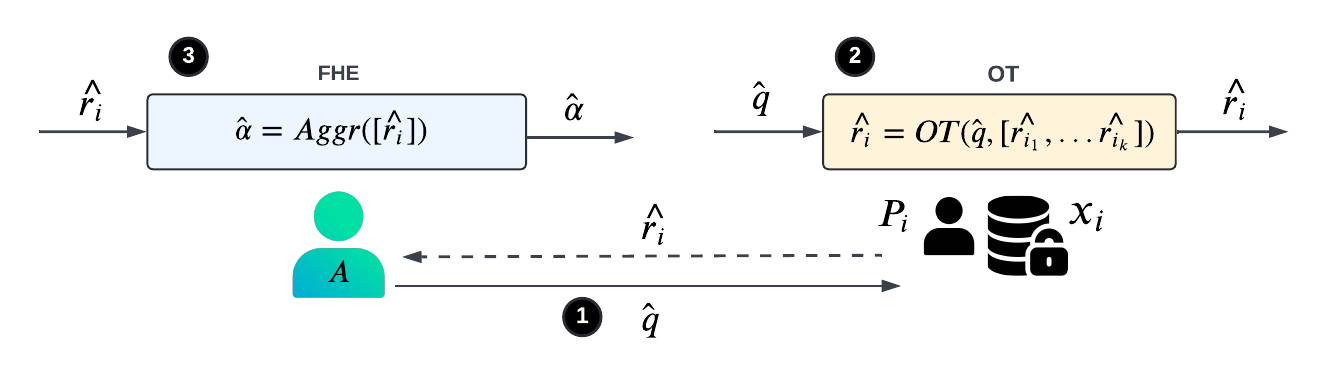}
		\caption{No TEE at P, no TEE at A, confidential query. The OT protocol involves multiple rounds of message exchanges and cryptographic operations between A and $P_i$. These details are omitted from the figure for simplicity.}
		\label{fig:arch-v5}
	\end{figure}

	Contrary to \vartwo, the query issued by $A$ to the data holders is confidential.
	This could be because it provides valuable insights or strategic information that could undermine the query issuer’s objectives.
	For instance, in our motivational example, the insurance company may want to keep private the specific disease it is inquiring about.
	Since there is no TEE at the parties (\autoref{fig:arch-v5}), a purely cryptographic technique is required to ensure the query is kept confidential.
	This can be achieved using a protocol like oblivious transfer.
	As previously outlined, OT enables a sender to transfer one of many pieces of information to a receiver without knowing which information was actually received by the receiver.
	To implement oblivious transfer, we maintain per party $P_i$, a set of $k$ possible query subresults:
$\{r_{i_1}, r_{i_2}, \ldots, r_{i_k}\}$, with corresponding ciphertexts:  $\{\hat{r_{i_1}}, \hat{r_{i_2}}, \ldots, \hat{r_{i_k}}\}$.
	The aggregator's query is part of the set of possible corresponding queries $q\in \{q_1, q_2, \ldots, q_k\}$.
	OT is used to obtain the encrypted subresult $\hat{r_{i_q}}$ corresponding to the query issued by $A$, without $P_i$ knowing which of the subresults was actually received by $A$, hence ensuring $P_i$ remains oblivious as to the contents of $q$.
	The encrypted subresults from all parties are then aggregated at $A$ via FHE and the aggregate decrypted via ThFHE.

	\subsubsection{No TEE at P, TEE at A, confidential query}
	\label{sec:defvar4}

	\begin{variantbox}{Variant 4: \( (P_\mathsf{NT}, A_\mathsf{T}, Q_\mathsf{C}) \)}
		\textbf{Protocol:}
		\begin{enumerate}[label=\arabic*]
			\item $pp=\setup()$
			\item $P_i$: $\{r_{i_{[k]}}\}=\evl\big(\{f_{q_{[k]}}\},x_i\big)$
			\item $P_i$: $\{\hat{r}_{i_{[k]}}\}=\pkenc\big(\{r_{i_{[k]}}\},pk_A\big)$
			\item $A,P_i$: $\hat{r}_{i}=\ot\big(q, \{\hat{r_{i_{[k]}}}\}\big)$
			\item $A$: $\{r_{[n]}\}=\teeskdec\big(\{\hat{r}_{[n]}\}, sk_A\big)$
			\item $A$: $\alpha=\teeagg\big(\{r_{[n]}\}\big)$
		\end{enumerate}
	\end{variantbox}

\begin{figure}[htbp!]
		\centering
		\includegraphics[width=0.4\textwidth]{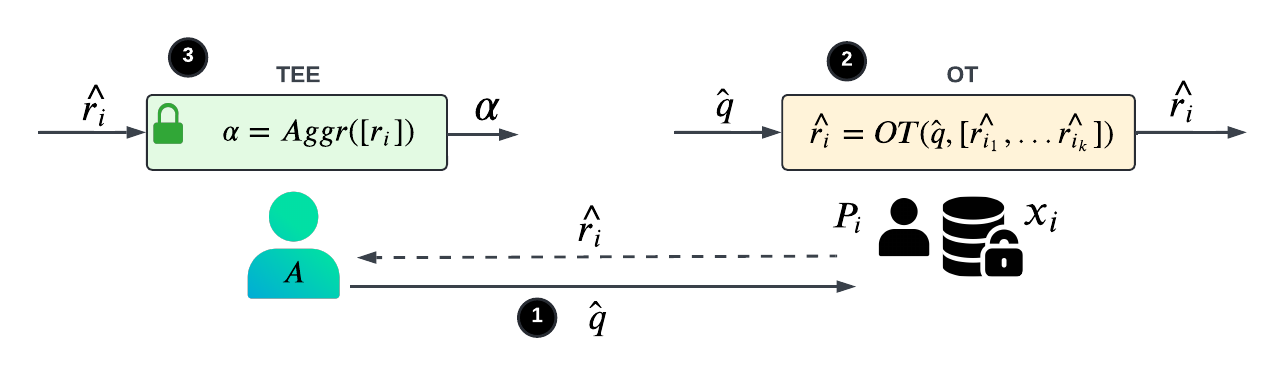}
		\caption{No TEE at P, TEE at A, confidential query}
		\label{fig:arch-v6}
	\end{figure}

	Similar to \varthree, the confidentiality requirement on the query coupled with the absence of TEEs at the parties means a purely cryptographic approach must be leveraged to ensure $q$ is kept confidential.
	OT is done as explained for \varthree to obtain the encrypted subresults obliviously at the aggregator.
	The subresults are encrypted by the parties using the aggregator's public key as shown in \autoref{fig:arch-v6}, and securely decrypted within the aggregator's TEE with its private key.

	\subsubsection{TEE at P, no TEE at A, confidential query}
	\label{sec:defvar5}

	\begin{variantbox}{Variant 5: \( (P_\mathsf{T}, A_\mathsf{NT}, Q_\mathsf{C}) \)}
		\textbf{Protocol:}
		\begin{enumerate}[label=\arabic*]
			\item $pp=\setup()$
			\item $A$: $\hat{q} =\pkenc(q,pk_{P_i})$
			\item $A$: $\snd_{A\rightarrow P_i}(\hat{q})$
			\item $P_i$: $q=\teeskdec(\hat{q}, sk_{P_i})$
			\item $P_i$: $r_i=\teeeval(f_q,x_i)$
			\item $P_i$: $\hat{r}_i = \tfheenc(r_i, pk)$
			\item $P_i$: $\snd_{P_i\rightarrow A}$($\hat{r_i}$)
			\item $A$: $\hat{\alpha} = \tfheeval(pk, f, \{\hat{r}_{[n]}\})$
			\item $A$: $\snd_{A\rightarrow P_i}$($\hat{\alpha}$)
			\item $P_i$: $\alpha_i = \tfhedec(\hat{\alpha}, {sk_{P_i}})$
			\item $P_i$: $\snd_{P_i\rightarrow A}$($\alpha_i$)
			\item $A$: $\alpha=\tfhecomb(\{\alpha_{[n]}\})$
		\end{enumerate}
	\end{variantbox}

\begin{figure}[t!]
		\centering
		\includegraphics[width=0.4\textwidth]{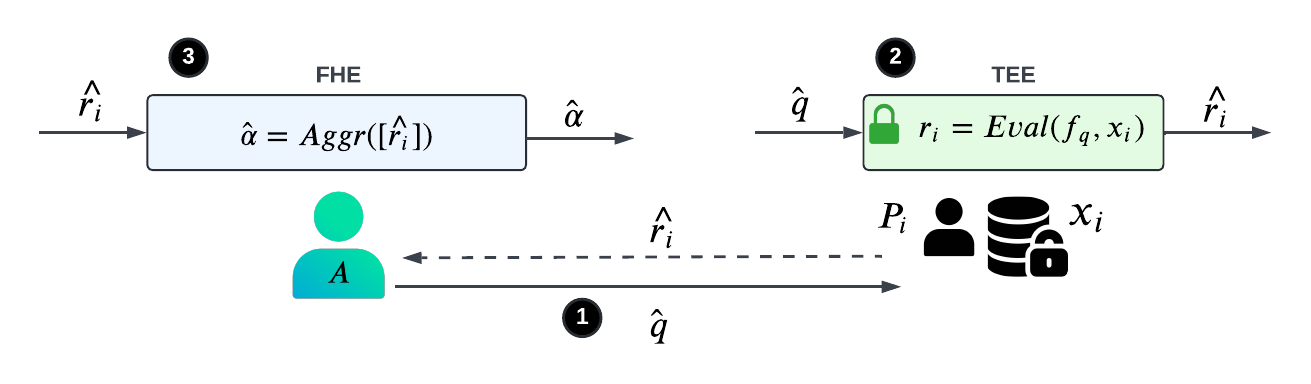}
		\caption{TEE at P, no TEE at A, confidential query}
		\label{fig:arch-v7}
	\end{figure}

	Here the aggregator first encrypts the query and sends it to the parties as shown in \autoref{fig:arch-v7}.
	Unlike \varfour, the presence of TEEs at the parties removes the need for a relatively expensive cryptographic approach like OT to ensure query confidentiality.
	So the query is simply decrypted and evaluated securely on the party's data from within a TEE.
	The subresult is encrypted inside the TEE and sent to the aggregator.
	The latter then aggregates all the subresults via FHE to obtain the encrypted aggregate which is decrypted via ThFHE.

	\subsubsection{TEE at P, TEE at A, confidential query}
	\label{sec:defvar6}

	\begin{variantbox}{Variant 6: \( (P_\mathsf{T}, A_\mathsf{T}, Q_\mathsf{C}) \)}
		\textbf{Protocol:}
		\begin{enumerate}[label=\arabic*]
			\item $pp=\setup()$
			\item $A$: $\hat{q}=\pkenc(q,pk_{P_i})$
			\item $A$: $\snd_{A\rightarrow P_i}(\hat{q})$
			\item $P_i$: $q =\teeskdec(\hat{q},sk_{P_i})$
			\item $P_i$: $r_i=\teeeval(f_q,x_i)$
			\item $P_i$: $\hat{r_i}=\pkenc(r_i,pk_A)$
			\item $P_i$: $\snd_{P_i\rightarrow A}(\hat{r_i})$
			\item $A$: $\{r_{[n]}\}=\teeskdec(\{\hat{r}_{[n]}\}, sk_A)$
			\item $A$: $\alpha=\teeagg(\{r_{[n]}\})$
		\end{enumerate}
	\end{variantbox}

\begin{figure}[htbp!]
		\centering
		\includegraphics[width=0.4\textwidth]{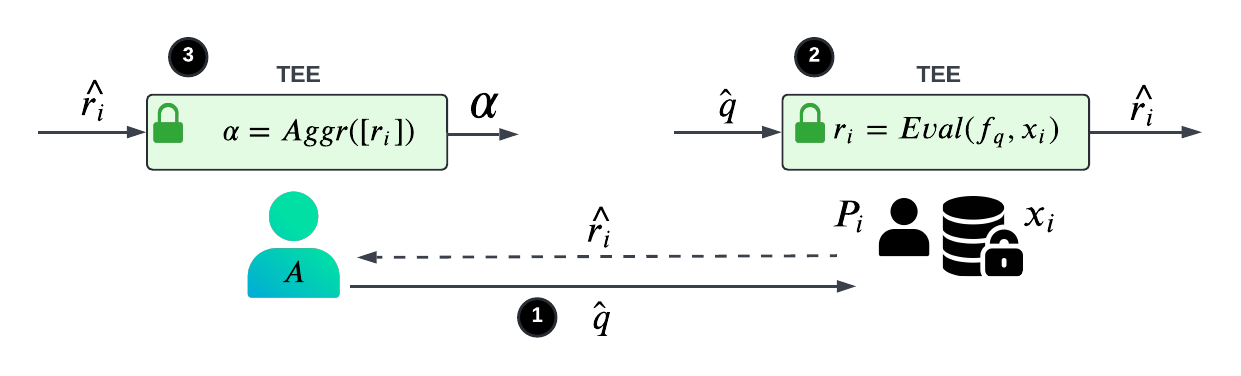}
		\caption{TEE at P, TEE at A, confidential query}
		\label{fig:arch-v8}
	\end{figure}

	Similar to \varfive, $A$ encrypts its confidential query and sends to the parties as shown in \autoref{fig:arch-v8}.
	Each party securely decrypts $q$ inside a TEE and executes it on their private data.
	The subresult is then encrypted within the TEE and sent back to $A$.
	Since the latter is equipped with a TEE, it securely decrypts all subresults from parties inside it's TEE (following \autoref{algo:tee-threshold}), and aggregates these subresults to obtain the final aggregate which is shared to all parties.

	\subsection{Heterogeneous data providers}
	The current SA variants explored assume a homogeneous setting where all data providers employ the same security mechanism--either TEEs or cryptographic techniques to protect query privacy. 
	A heterogeneous architecture, where data providers adopt different security approaches, can be constructed by simply combining existing variants.
	For instance, combining \varsix with \varfour or \varfive with \varthree enables scenarios where some data provid\-ers use TEEs while others use a purely cryptographic technique for query privacy, with an aggregator which uses a TEE or a purely cryptographic approach for privacy of subresults.

	\subsection{Discussion}
	\subpoint{Malicious aggregator.}
	In the considered adversary model, all parties, including the aggregator, are assumed to be semi-honest, and thus follow the protocol faithfully.
	This assumption precludes scenarios where a malicious aggregator might compute the final aggregate based on a single subresult, which could indirectly reveal a party's private information.
	In many real-world settings, such an assumption may not suffice.
	Thus, we discuss the security implications of the presence of such a malicious aggregator, and propose measures to improve the system's robustness.

	For models with a TEE-enabled aggregator, such attacks are mitigated by default through the remote attestation protocol, which allows all parties to cryptographically verify that the aggregator's TEE is running the agreed-upon threshold aggregation function, whose implementation ensures aggregation only after the given threshold of subresults has been received (\autoref{algo:tee-threshold}).
	In models using ThFHE, the latter provides cryptographic guarantees that the aggregation is computed on the correct number of subresults.
	Otherwise, in the absence of a TEE on an aggregator and ThFHE, the parties could proceed to re-compute the aggregate value by splitting their subresults into $n$ shares and employing a traditional secret sharing scheme~\cite{danezis13}.
	The result can then be used to validate or verify the aggregate computed by the aggregator.
	Such verification mechanisms can address scenarios where a malicious aggregator might compute aggregates using fewer than the required threshold of subresults.
	
	\subpoint{Malicious data providers.}
	In general, preventing attacks on the protocol by a malicious data provider is more challenging.
	For example, a malicious data provider could submit false data, leading to an inaccurate final aggregate.
	This issue can be mitigated by requiring each data provider to share cryptographic signatures of their private data beforehand.
	These signatures could then be used later to verify the validity of the subresults provided by the party.
	Designs that keep the query confidential also reduce the likelihood of such attacks, as the data provider has limited knowledge of how their data is being evaluated.
	Additionally, systems could be implemented to incentivize data providers to submit accurate data.

\section{Implementation}
\label{sec:implem}

\smallskip\noindent\textbf{Secure aggregation variant implementations.}
Some SA variants we explore share the same code implementation, the only difference being whether the code is executed within a TEE or not.
We have four distinct code categories in total which can be derived by combining the possibilities for input privacy and query confidentiality as follows:
Input privacy ($\mathsf{IP}$), \ie confidentiality of each party's data, is ensured either via a TEE at the aggregator ($\mathsf{IP}_\mathsf{TEE}$) or via FHE ($\mathsf{IP}_\mathsf{FHE}$).
Similarly, query confidentiality ($\mathsf{QC}$) can be achieved using either TEEs at the parties ($\mathsf{QC}_\mathsf{TEE}$) or oblivious transfer ($\mathsf{QC}_\mathsf{OT}$).
As such, the code implementations of all the SC variants explored can be categorized into six categories: $\{\mathsf{IP}_\mathsf{TEE},\mathsf{IP}_\mathsf{FHE}\}\times\{\mathsf{QC}_\mathsf{TEE},\mathsf{QC}_\mathsf{OT}, \mathsf{QNC}\}$.
The code implementation for $\mathsf{QC}_\mathsf{TEE}$ and $\mathsf{QNC}$ are the same, the only difference being the former runs within a TEE and the latter without.
We refer to this common code implementation as $\mathsf{Q}_\mathsf{comm}$.
This leaves us with four unique code categories encompassing all the architectures studied.
We classify each SA variant into these categories in \autoref{tab:code-fam}.

\begin{table}[!t]
	\centering
	\small
	\caption{SA variants explored and their corresponding code categories.}
	\label{tab:code-fam}
	\setlength{\tabcolsep}{2pt}
	\rowcolors{1}{gray!0}{gray!10} 
		\begin{tabular}{lll}
			\toprule
			\rowcolor{gray!25}
			\textbf{Code category}                                & & \textbf{SA variants}      \\
			\midrule
			$\mathsf{IP}_\mathsf{TEE},\mathsf{QC}_\mathsf{comm}$: & \texttt{iptee-qccom} & \vartwo, \varsix  \\
			$\mathsf{IP}_\mathsf{TEE},\mathsf{QC}_\mathsf{OT}$:   & \texttt{iptee-qcot}    & \varfour          \\
			$\mathsf{IP}_\mathsf{FHE},\mathsf{QC}_\mathsf{comm}$: & \texttt{ipfhe-qccom} & \varone, \varfive \\
			$\mathsf{IP}_\mathsf{FHE},\mathsf{QC}_\mathsf{OT}$:   & \texttt{ipfhe-qcot}    & \varthree         \\
			\bottomrule
			\rowcolor{white}
		\end{tabular}
\end{table}

We implemented all FHE algorithms on top of OpenFHE~\cite{openfhe}, a popular FHE library which provides ThFHE.
The TEEs at the aggregator and parties were implemented as Intel SGX enclaves.
The library OS Occlum~\cite{occlum} was used to run legacy code inside the SGX enclaves.
The oblivious transfer protocol was implemented with \text{libOTe}~\cite{libote}, a fast and portable C++20 library for OT. We used the KKRT protocol~\cite{kkrt-ot}.

\section{Evaluation}
\label{sec:eval}
Our experimental evaluations seek to answer the following questions:
\begin{itemize}[]
	\item[\textbf{Q1}:] How do computational and communication overhead vary across the secure computation variants?
	\item[\textbf{Q2}:]  How do TEE and purely cryptographic security techniques impact the performance of the SA variants?
	\item[\textbf{Q3}:] What is the impact of query confidentiality on the performance of the SA variants?
	\item[\textbf{Q4}:]  How does the memory overhead vary across the SA variants?

\end{itemize}
\subpoint{Methodology.}
We run the aggregator and parties as different processes on the same server, and measure the end-to-end completion time of the SA protocols.
Each data provider's database is represented as an integer vector of arbitrary size.
We show the cost of important online phases such as query execution, aggregation, and communication, and study the effects of the different security techniques leveraged on the performance of the SA protocol.

\subpoint{Server setup.}
Our evaluations are conducted on a server equipped with an 8-core Intel Xeon Gold 5515+ processor clocked at 3.20 GHz, a 22.5 MB last-level cache, and 128GB of DRAM.
The server runs Ubuntu 22.04.4 LTS and Linux 5.15.0-122-generic.
We use Occlum containers based on version 0.31.0-rc-ubuntu22.04.
The configured EPC size is 64GB.
We report the median of 10 complete runs of the protocol for each data point.
For all plots, K=$\times1000$ and M=$\times1000000$.

\subsection{Performance analysis of SA variants}
\begin{figure}[!t]
	\centering
	\includegraphics[scale=0.9]{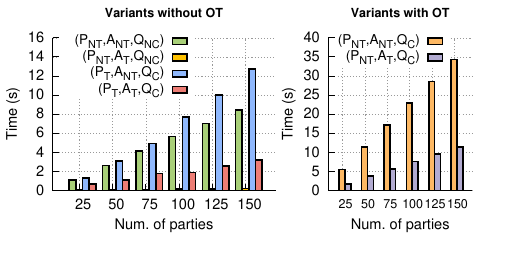}
	\caption{Execution time for the different SA variants with a varying number of parties each with DB size of 10000 elements.}
	\label{fig:varnode}
\end{figure}

\begin{figure}[!t]
	\centering
	\includegraphics[scale=0.9]{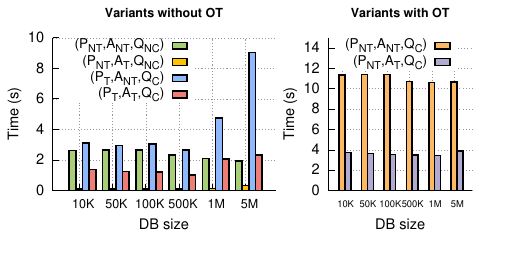}
	\caption{Execution time for the different SA variants with 50 parties and varying dataset sizes.}
	\label{fig:vardb}
\end{figure}

\begin{figure}[!t]
	\centering
	\includegraphics[scale=0.9]{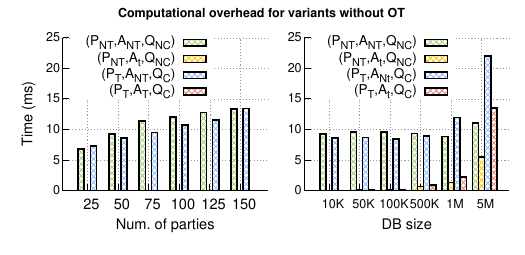}
	\caption{Computational overhead of different variants.}
	\label{fig:computation}
\end{figure}
We evaluate the performance of the different variants following two metrics: total computation costs and execution time.
The computation costs include encryption, decryption, query evaluation, and aggregation overhead, while the execution time represents the execution time from the time the request is sent by the aggregator to the time the final result is obtained. We can deduce the communication costs including all data exchange overhead, \eg sending queries to parties and obtaining the subresults, by subtracting the computation costs from the execution time.
We distinguish between variants which use oblivious transfer to evaluate the query and those which don't.
For the variants that do not utilize OT, the total time for the protocol is divided into computation and communication costs, while for the variants with OT, only the total execution time is reported, as computation and data exchange are intertwined in the OT protocol.
We analyze the impact of the chosen security techniques and query confidentiality on the performance of the SA variants.

\autoref{fig:varnode} and \autoref{fig:vardb} show the combined performance (includes both computational and communication overhead) of all variants while varying the number of parties and database (\ie data share) sizes respectively.

\smallskip\noindent\textbf{Computational and communication overhead of SA variants:} \emph{Answer to Q1 and Q2.}

\autoref{fig:varnode}, \autoref{fig:vardb}, and \autoref{fig:computation} show that communication overhead dominates the overall cost in all variants; this is coherent with previous work ~\cite{piranha}.
Notably, when comparing crypto-only variants (\eg \varone) to TEE-only variants (\eg \vartwo), the TEE-based approach reduces communication overhead due to the smaller size of the partial results: a single integer value for TEE-based variants versus an FHE-generated ciphertext, which is a structure of size $386KB$ (almost $100$K$\times$ larger than a $4B$ integer).
Moreover, with FHE, there are more message exchanges between the aggregator and the parties, unlike the TEE-based variant, where only the partial results are sent to the aggregator.
This difference in communication overhead is evident when comparing \varone, where FHE is employed for computing partial and final results, to \vartwo, where a TEE is used.
Here, the TEE-based approach achieves a performance improvement of about $41.46\times$.
The communication overhead scales linearly with the number of parties, mainly due to the increased number of TCP connections between the aggregator and the parties.
In real-world applications involving communications over a wide area network (WAN), the communication overhead is expected to be slightly larger.

\observ{The communication overhead dominates the overall overhead in all SA variants; using TEEs at parties improves the communication overhead by up to $41\times$ with respect to FHE.}

\smallskip\noindent\textbf{Impact of security techniques on computational performance:} \emph{Answer to Q2.}

As illustrated in \autoref{fig:computation}, using TEEs results in significantly lower computational costs when compared to purely cryptographic techniques.
This can be observed by comparing the variants where only the security mechanism differs at either the aggregator or the parties.
For example, when varying the number of parties while keeping the DB size constant for all parties, \varone is up to $785\times$ slower compared to \vartwo, and \varfive is up to $583\times$ slower compared to \varsix.
A similar trend is observed when considering the variants with OT: \varthree and \varfour.
This substantial performance drop is primarily due to the use of FHE at the aggregator in \varone and \varfive, as opposed to a TEE.
The reduced cost with TEEs can be attributed to their ability to securely process unencrypted (\ie decrypted) data directly in the CPU, unlike FHE which performs computations on encrypted data.

\observ{Using a TEE at the aggregator as opposed to FHE improves computational overhead by up to $785\times$.}

\smallskip\noindent\textbf{Impact of query confidentiality on performance:} \emph{Answer to Q3.}

The impact of query confidentiality on the overall performance of SA variants can be observed by comparing variants that differ only in their approach to query confidentiality.
This involves the variants with oblivious transfer: \varthree and \varfour (which ensure query confidentiality), and their non-OT counterparts: \varone and \vartwo (which do not ensure query confidentiality).
Specifically, when varying the number of parties while keeping the DB size constant (\ie \autoref{fig:varnode}), \varthree is up to $4\times$ slower when compared to \varone, and \varfour is up to $56\times$ slower when compared to \vartwo.
This significant performance drop is primarily attributed to the large cryptographic overhead introduced by the oblivious transfer protocol required for confidential queries in the absence of TEEs at the parties.

\observ{In the absence of TEEs at the parties, query confidentiality introduces up to $56\times$ overhead due to the high cost of oblivious transfer.}

\renewcommand{\arraystretch}{1.2}
\begin{table*}[!ht]
	\centering
	\small
	\caption{Table summarizing the normalized overheads of the SA variants relative to a non-secure variant, how scalable each variant is, and the security guarantees provided for parties' input data ($x_i$) and the aggregator's query ($q$). The baseline represents a scenario without security guarantees. Lower normalized overheads indicate better overall performance, while smaller scalability gradients reflect improved scalability. The negative scalability values observed are primarily due to measurement variations in communication overhead between parties and the aggregator, and do not represent a consistent trend. The overheads reported are computed for 150 parties and a DB size of 10000 values per party.}
	\label{tab:summary}
	\rowcolors{1}{gray!0}{gray!10} 
		\begin{tabular}{lllllllll} 
			\toprule
			\rowcolor{gray!25}
                \multirow{2}{*}{} & \multicolumn{2}{c}{Performance} & \multicolumn{2}{c}{Memory (aggregator + parties)} & \multicolumn{2}{c}{Scalability} & \multicolumn{2}{c}{Confidentiality} \\
                \rowcolor{gray!25}
			Variant & Total cost (s) & Norm. overhead & Total memory (GB) & Norm. overhead & Parties & DB size & $x_i$ & $q$ \\
			\midrule
                Baseline  & 1.94e-1  & $1\times$ & 3.47e-1 & $1\times$ & 6.72e-4 & 1.81e-6 & \no & \no \\
			\varone  & 8.43  & $43.474\times$ & 7.73 & $22.3\times$ & 5.85e-2 & -1.27e-7 & \yes & \no \\
			\vartwo   & 2.03e-1  & $1.046\times$ & 3.47e-1 & $1\times$ & 1.13e-3 & 4.87e-8 & \yes & \no \\
			\varthree   & 34.3  & $176.793\times$ & 14.1 & $40.689\times$ & 2.30e-1 & -1.22e-7 & \yes & \yes \\
			\varfour   & 11.4  & $58.861\times$ & 4.13e-1 & $1.19\times$ & 7.73e-2 & 5.04e-8 & \yes & \yes \\
			\varfive   & 12.7  & $65.5\times$ & 7.73 & $22.3\times$ & 9.19e-2 & 1.24e-6 & \yes & \yes \\
                \varsix   & 3.22  & $16.577\times$ & 3.47e-1 & $1\times$ & 1.94e-2 & 2.19e-7 & \yes & \yes \\
			\bottomrule
			\rowcolor{white}
		\end{tabular}
\end{table*}

\subsection{Memory overhead}
\begin{figure}[ht!]
	\centering
	\includegraphics[scale=0.9]{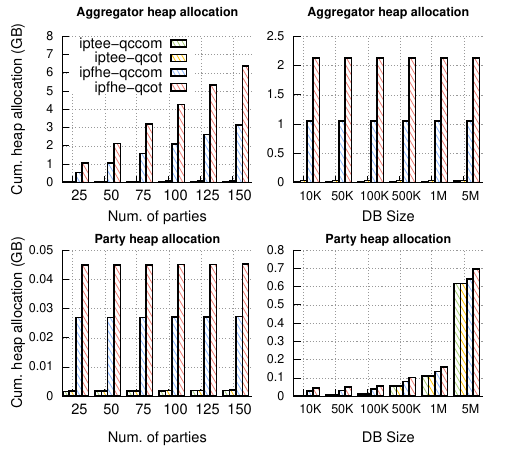}
	\caption{Cumulative heap allocation for parties and aggregator for a varying number of parties and DB sizes.}
	\label{fig:memory}
\end{figure}
In this section, we evaluate the memory overhead, measured as \emph{cumulative heap allocation} (\ie the total memory allocated from the heap during a program's execution) of the SA variants.
We recall from \S\ref{sec:implem} that the SA variants can be categorized into four code categories: \texttt{iptee-qcomm}, \texttt{iptee-qcot}, \texttt{ipfhe-qccom}, and \texttt{ipfhe-qcot}.
Using the gperftools library~\cite{gperf}, we tracked the cumulative heap allocation at both the aggregator and the parties for the execution of each code category.
\autoref{fig:memory} shows the results obtained, and the key findings are summarized below.

As the number of parties increases, the cumulative memory usage on the aggregator increases since it has to retrieve more partial results from parties.
Moreover, when a cryptographic technique, \ie FHE, is employed at the aggregator for input privacy, \eg \texttt{ipfhe-qccom}, the memory overhead at the aggregator increases up to  $22.28\times$ compared to when a TEE is used, \eg \texttt{iptee-qccom}.
This is mainly due to the larger size of ciphertexts in FHE compared to the TEE approach.
In a purely cryptographic scenario like \texttt{ipfhe-qcot}, where FHE ensures input privacy at the aggregator and OT ensures query confidentiality, the memory overhead at the aggregator is up to $40.69\times$ higher than in a TEE-only scenario like \texttt{iptee-qccom}.
Lastly, the memory overhead at the parties increases linearly with the DB size, which is expected.
However, an increase in DB size per party has no impact on the memory overhead at the aggregator, provided the number of parties remains constant. 
This is because the aggregator receives and processes the same number of ciphertexts (\ie partial results) regardless of the DB size per party.

\observ{
Using TEEs for secure aggregation reduces memory overhead by up to $40\times$ compared to purely cryptographic techniques like FHE and oblivious transfer.
}

\subsection{Scalability analysis}
To assess the scalability of each variant, we employ a straightforward linear regression approach to determine the gradient/slope of the respective graph as the number of parties and database sizes increase. 
In the context of our work, we refer to this metric as the \emph{scalability gradient}, which quantifies how rapidly the total cost grows with an increasing number of parties or DB size. 
The results are summarized in \autoref{tab:summary}.
Lower scalability gradients represent better scalability, signifying that the SA variant incurs minimal overhead as the workload increases.
Our analysis reveals that the TEE-based approaches generally exhibit better scalability compared to the crypto-based approaches. 
For example, the scalability gradient of \vartwo is about $51.77\times$ smaller than that of \varone, demonstrating that the cost of the SA protocol increases $51.77\times$ more gradually when a TEE is used at the aggregator instead of FHE.
Any decreases in communication overhead with larger DB size (\ie \autoref{fig:vardb}) are mainly due to variations in the measurements of communication overhead between the aggregator and the parties, and do not represent a consistent trend.
Conversely, computational costs remain relatively stable for database sizes.
The computations performed at the parties primarily involve basic operations such as sums, multiplications and counts, which contribute to minimal processing overhead in general.

\observ{
TEEs provide improved scalability with respect to purely cryptographic approaches.
}

\subsection{Security considerations}
Despite the improved performance of TEEs compared to cryptographic approaches, they are prone to certain hardware-specific vulnerabilities~\cite{kocher2020spectre, meltdown, sgx-spectre, plundervolt} and require trusting the processor manufacturer.
Cryptographic methods like FHE and OT, by contrast, provide stronger, mathematically backed security guarantees.
The choice between these techniques at the aggregator or parties may depend on factors such as regulatory constraints, performance needs, and hardware availability, such as the presence of Intel SGX at the parties or aggregator.
In less trusted environments, cryptographic approaches may be preferable for stronger security.
Conversely, in performance-critical scenarios, TEEs might take precedence.
For example, if parties are concerned about leakage of sensitive data (from multiple parties) at the aggregator via TEE-based side-channel attacks, they may prefer to adopt an architecture where a cryptographic approach is used at the aggregator.
The aggregator on the other hand may accept TEE techniques at the parties for maintaining query confidentiality. 
Such a scenario will correspond to variant \varfive defined in page \pageref{sec:defvar5}.
Similarly, if TEE hardware is unavailable at some parties, a purely cryptographic approach could be adopted for the parties by choosing a variant like \varfour defined in page \pageref{sec:defvar4}.

\observ{
The decision on which variant to adopt hinges on a combination of factors, including regulatory compliance, system performance requirements, hardware availability (such as the presence of TEE-enabled devices), potential attack vectors (\eg feasibility of TEE side-channel attacks), and the specific security objectives of the deployment environment. 
}

\section{Related work}
\label{sec:rw}
In this section, we explore related work under the following categories:
\emph{(i)}~Secure aggregation with purely software-based privacy techniques,
\emph{(ii)}~TEE-assisted secure aggregation, and 
\emph{(iii)}~Hybrid secure aggregation approaches, 
and contrast them with the approaches we propose.

\subpoint{Secure aggregation with purely software based privacy techniques.}
Several studies have proposed different approaches to secure aggregation, including differential privacy~\cite{shi2011privacy, roth19, tariq14}, homomorphic encryption~\cite{hiccups, hosseini21}, lattice cryptography~\cite{bell23} and secret sharing~\cite{bonawitz2017practical, stevens2022secret}.
Rather than rely solely on cryptographic techniques or differential privacy, our work explores various ways to integrate TEEs into secure aggregation architectures, achieving strong privacy guarantees and significantly reduced computational costs.

\subpoint{TEE-assisted secure aggregation.}
Several studies have employed hardware-based TEEs for secure aggregation, particularly in federated learning. 
Notable examples include~\cite{zhao22, olga16, zhang21}, which utilize TEEs to ensure privacy of machine learning gradients from data providers.
While these work highlight the effectiveness of TEEs in secure aggregation, they explore only very specific secure aggregation architectures.
For example, \cite{olga16} and \cite{zhang21} correspond to variant \vartwo of our work, where all parties send their data to a central aggregator equipped with a TEE.
Our work broadens the scope by thoroughly exploring multiple SA architectures, analyzing their trade-offs in security and performance.

\subpoint{Hybrid SA approaches.~}
Relatively fewer work have explored secure computation architectures combining TEEs and cryptographic techniques.
\cite{wu2022hybrid} presents a hybrid MPC scenario where there are varying degrees of trust perceived by parties in TEEs, and leverage the latter selectively for parts of their software, and cryptographic techniques for the rest.
\cite{choncholas2023tgh} proposes a hybrid protocol for FaaS platforms where computations are moved from SGX enclaves to garbled circuits to address memory constraints in SGX enclaves.
Similarly, \cite{dantonio2023} utilizes TEEs to compute expensive homomorphic encryption operations like noise refreshing.
These work are orthogonal to ours in that they aim to mitigate the cost of cryptographic approaches by outsourcing some computations to TEEs. 

\section{Conclusion}
\label{sec:conclusion}
This paper explores how TEEs and purely cryptographic security techniques can be combined in secure aggregation architectures, and discusses the associated performance and security trade-offs. 
Our evaluations demonstrate that TEEs can enhance both communication and computation performance overhead in SA variants, achieving improvements of up to $41\times$ and $785\times$, respectively, when compared to purely cryptographic techniques like FHE.

\smallskip\noindent\textbf{Applications of the SA variants.}
The SA architectures presented have numerous real-world applications, \eg in healthcare research for privacy-preserving disease tracking across healthcare providers~\cite{buchanan2020review, reichert2020privacy}, e-healthcare data aggregation~\cite{tang-ehealth}, privacy-preserving auctions~\cite{bogetoft2009secure, aly17}, privacy preserving aggregation in smart grids~\cite{danezis13, garcia2011privacy}, secure e-voting~\cite{kusters2020ordinos}, privacy preserving federated learning~\cite{wen2023} and privacy preserving opinion aggregation~\cite{kiayias2022privacy}, among others.

\section*{Acknowledgment}
This work was supported by the Swiss National Science Foundation under project P4: Practical Privacy-Preserving Processing (no. 215216).
\bibliographystyle{ACM-Reference-Format}
\bibliography{references,dblp}

\end{document}